\documentclass[twocolumn]{aastex63}
\usepackage{apjfonts}
\usepackage{graphicx}
\usepackage{xspace}
\usepackage{comment}    

\begin{document}

\lefthead{Chandra Observations of IGR Sources}
\righthead{Tomsick et al.}

\def\lsim{\mathrel{\lower .85ex\hbox{\rlap{$\sim$}\raise
.95ex\hbox{$<$} }}}
\def\gsim{\mathrel{\lower .80ex\hbox{\rlap{$\sim$}\raise
.90ex\hbox{$>$} }}}

\title{Using Chandra Localizations and Gaia Distances and Proper Motions to Classify Hard X-ray Sources Discovered by INTEGRAL}

\author{John A. Tomsick}
\affiliation{Space Sciences Laboratory, 7 Gauss Way, University of California, Berkeley, CA 94720-7450, USA}

\author{Benjamin M. Coughenour}
\affiliation{Space Sciences Laboratory, 7 Gauss Way, University of California, Berkeley, CA 94720-7450, USA}

\author{Jeremy Hare}
\affiliation{NASA Goddard Space Flight Center, Greenbelt, MD 20771, USA}
\affiliation{NASA Postdoctoral Program Fellow}

\author{Roman Krivonos}
\affiliation{Space Research Institute, Russian Academy of Sciences, Profsoyuznaya 84/32, 117997 Moscow, Russia}

\author{Arash Bodaghee}
\affiliation{Georgia College \& State University, CBX 82, Milledgeville, GA 31061, USA}

\author{Sylvain Chaty}
\affiliation{Universit\'e de Paris and Universit\'e Paris Saclay, CEA, CNRS, AIM, F-91190 Gif-sur-Yvette, France}
\affiliation{Universit\'e de Paris, CNRS, AstroParticule et Cosmologie, F-75013 Paris, France}

\author{Ma\"ica Clavel}
\affiliation{Univ. Grenoble Alpes, CNRS, IPAG, F-38000 Grenoble, France}

\author{Francesca M. Fornasini}
\affiliation{Stonehill College, 320 Washington Street, Easton, MA 02357, USA}

\author{Jerome Rodriguez}
\affiliation{D\'ept. of Astrophysics,  lab AIM, CEA/IRFU, CNRS/INSU, Universit\'e Paris-Saclay,  Universit\'e de Paris, Orme des Merisiers F-91191, Gif-sur-Yvette, France}

\author{Aarran W. Shaw}
\affiliation{Department of Physics, University of Nevada, Reno, NV 89557, USA}

\begin{abstract}

Here we report on X-ray observations of ten 17--60\,keV sources discovered by the INTEGRAL satellite.  The primary new information is sub-arcsecond positions obtained by the Chandra X-ray Observatory.  In six cases (IGR~J17040--4305, IGR~J18017--3542, IGR~J18112--2641, IGR~J18434--0508, IGR~J19504+3318, and IGR~J20084+3221), a unique Chandra counterpart is identified with a high degree of certainty, and for five of these sources (all but J19504), Gaia distances or proper motions indicate that they are Galactic sources.  For four of these, the most likely classifications are that the sources are magnetic Cataclysmic Variables (CVs).  J20084 could be either a magnetic CV or a High Mass X-ray Binary.  We classify the sixth source (J19504) as a likely Active Galactic Nucleus (AGN).  In addition, we find likely Chandra counterparts to IGR~J18010--3045 and IGR~J19577+3339, and the latter is a bright radio source and probable AGN.  The other two sources, IGR~J12529--6351 and IGR~J18013--3222 do not have likely Chandra counterparts, indicating that they are transient, highly variable, or highly absorbed.

\end{abstract}


\section{Introduction}

The International Gamma-Ray Astrophysics Laboratory \citep[INTEGRAL][]{winkler03}, which launched in 2002, has uncovered a large number of new or previously poorly studied sources by surveying the sky and Galaxy in the $\sim$20-100\,keV band.  These are called ``INTEGRAL Gamma-Ray'' (IGR) sources.  To produce emission in this energy band requires particles to be accelerated to high energies, and most of the IGR sources are places where extreme physics is taking place.  In many cases, the extreme physics is related to accretion onto compact objects (black holes, neutron stars, or magnetic white dwarfs) or to highly-magnetized neutron stars (pulsars or magnetars).  

An INTEGRAL catalog based on 8 years of observations includes 939 persistent and transient sources detected above the 4.5$\sigma$ significance level in the 17--100\,keV band \citep{bird16}. The largest groups of sources include 369 Active Galactic Nuclei (AGN), 129 Low Mass X-ray Binaries (LMXBs), 116 High Mass X-ray Binaries (HMXBs), and 56 Cataclysmic Variables (CVs).  Of the 939 sources, the source type was unknown for 219 at the time \cite{bird16} was written.  The 939 sources include both previously known sources and 447 IGR sources.  The main reasons that INTEGRAL has found new (IGR) sources are: 1. the relatively large field of view has allowed for the full sky to be covered; 2. the high-energy bandpass is not affected by Galactic absorption; and 3. most high-energy sources are transient or variable.  While there are many notable sources among the IGR sources, INTEGRAL's capabilities have been especially good for finding new HMXBs, including a population of obscured HMXBs \citep{mg03,walter06} and a population of Supergiant Fast X-ray Transients \citep{negueruela06,sguera06,romano14}.  

While the \cite{bird16} analysis was carried out for INTEGRAL observations of the whole sky and included enhancements for finding transient sources, another INTEGRAL analysis effort has focused on sources within $17.5^{\circ}$ of the Galactic plane \citep{krivonos12}.  The \cite{krivonos12} 9-year survey resulted in detections of 402 sources.  While these included persistent and transient sources, the search focused on sources that were detected in the combined 9-years of observations, which favors the detection of persistent sources.  The most recent report on this survey used 14 years of INTEGRAL data \citep{krivonos17}, and we selected sources from this catalog for the current work.  

The current work is focused on making progress toward classifying IGR sources of currently unknown nature by observing a selection of them with the Chandra X-ray Observatory.  INTEGRAL provides detections, high-energy spectra, and localization with 90\% confidence uncertainties of $1^{\prime}$--$5^{\prime}$, depending on source strength.  With such positional uncertainties, it is not usually possible to identify an optical or near-IR counterpart, especially in the Galactic plane.  Thus, the most important information that Chandra can provide is a more precise source position.  In addition, with its 0.3--10\,keV coverage, Chandra provides information about the soft X-ray energy spectrum.  We have carried out similar studies in the past \citep[e.g.,][]{tomsick12a,tomsick16}, and many of the analysis techniques that we employ in this work are the same as described in \cite{tomsick20}.  

\subsection{Target Selection}

We selected targets to observe with Chandra from the 72 new sources detected in the 17--60\,keV band and reported in the 14-year INTEGRAL catalog \citep{krivonos17}.  This catalog includes only sources that were not detected in the previous catalog version \citep{krivonos12}.  We eliminated sources with likely or definite classifications and with previous coverage with Chandra, XMM-Newton, or Swift.  As our scientific focus is on Galactic sources, we also used Galactic latitude as a criterion, and |$b$| is between $0.23^{\circ}$ and $6.27^{\circ}$ for the sources we study in this work.  We obtained Chandra observations for eight sources in Chandra cycle 20 (observations carried out in 2019) and five sources during cycle 19.  Results from three of the cycle 19 observations are reported in \cite{hare19} and Hare et al., submitted to ApJ, providing classifications of one HMXB and two magnetic CVs.  We report the results for the Chandra observations for the other two sources in this work.

In summary, we are reporting on short ($\sim$5\,ks) Chandra observations of the 10 IGR sources listed in Table~\ref{tab:krivonos}.  They are listed in order of how close they are to a Galactic latitude of $b = 0^{\circ}$.  In addition to the INTEGRAL source coordinates, we provide the flux and detection significance from \cite{krivonos17}.  We note that \cite{krivonos17} does not provide position uncertainties for individual sources but indicates that the 90\% confidence INTEGRAL error radii are typically $3.6^{\prime}$.  In Section 2, we describe the Chandra observations, analysis, and results, including searching for Chandra counterparts to the IGR sources, carrying out Chandra photometry, and providing X-ray localizations.  Section 3 includes the results of fitting Chandra and INTEGRAL energy spectra.  In Section 4, we search for multi-wavelength counterparts using the accurate Chandra localizations.  Section 5 and 6 include discussion of the results and conclusions, respectively.

\section{Chandra Observations, Analysis, and Results}

Table~\ref{tab:obs} provides the basic information for the ten Chandra observations, which occurred between 2018 February 20 and 2019 May 29.  We used the ACIS-I instrument \citep{garmire03} with exposure times of $\sim$5000\,s, which is sufficient to expect $>$100 counts based on the INTEGRAL fluxes and extrapolation of a hard power-law into the Chandra bandpass.  The field of view of the four ACIS-I chips is $16.9^{\prime} \times 16.9^{\prime}$, which easily contains the 90\% confidence INTEGRAL error regions with one pointing per source.  In each case, the pointing positions are the INTEGRAL positions given in Table~\ref{tab:krivonos}.

We reduced the data using the Chandra Interactive Analysis of Observations \citep[CIAO,][]{fruscione06} version 4.13 software with the Calibration Database (CALDB) version 4.9.4, largely following instructions in the CIAO science threads\footnote{See http://asc.harvard.edu/ciao/threads/index.html}.  We used {\ttfamily chandra\_repro} for reprocessing the data, resulting in a photon list along with other instrument files.  

We created exposure-corrected images in the 0.3--10 keV band for each of the 10 Chandra observations using {\ttfamily fluximage}. These images were then combed for sources by applying the wave detection algorithm {\ttfamily wavdetect}, using wavelets with scales of 1, 2, 4, 6, 8, 12, 16, 24, and 32, and with a detection threshold set to produce just one spurious source among the detection results. The number of sources detected for each observation is given in Table~\ref{tab:shifts}.

In order to account for any systematic uncertainty in our Chandra astrometry due to telescope pointing, we compared the detected source positions with several optical and near-IR catalogs using {\ttfamily wcs\_match}, and shifted the source positions for each ObsID accordingly. The catalogs we used to match our detected Chandra sources include Gaia EDR3 \citep{gaia_edr3}, the Variables in the Via Lactea (VVV) Survey by the Visual and Infrared Survey Telescope for Astronomy \citep[VISTA;][]{minniti10, mcmahon13, minniti17}, the UKIRT Infrared Deep Sky Survey \citep[UKIDSS;][]{lucas08}, and the Panoramic Survey Telescope and Rapid Response System \citep[PanSTARRS;][]{chambers16}. For each observation we chose the survey that provided the most cross-matches with our Chandra sources, while ignoring the candidate IGR counterpart (when applicable, see Section~2.1 for details on counterpart selection), and these results are provided in Table~\ref{tab:shifts}. We also include the average and maximum residuals between the Chandra and optical/near-IR catalog positions, after translation of the Chandra coordinates. In observations where fewer than 3 cross-matches were found, as was the case for ObsIDs 21248, 21250, 21251, and 21252, we did not shift the Chandra positions.

We performed Chandra aperture photometry to determine the number of counts for all of the detected sources.  We made a point spread function (PSF) map using {\ttfamily mkpsfmap}\footnote{We later realized that {\ttfamily fluximage} can provide this same map.} for an energy of 2.3\,keV (the typical average photon energy for the full 0.3--10\,keV Chandra bandpass), and determined the 95\% encircled energy radius for each source.  After defining background regions for each observation, we used {\ttfamily dmextract} to extract background-subtracted counts in the 0.3--2\,keV, 2--10\,keV, and 0.3--10\,keV energy bands.  

\subsection{Selecting the Most Likely Counterparts}

As in previous work \citep[e.g.,][]{tomsick20}, we calculated the probability that sources with the brightnesses we observe would be found in a search area with a radius of $\theta_{\rm search}$ by chance.  In cases where the source is within the 90\% confidence INTEGRAL error radius ($\theta_{INTEGRAL}$), $\theta_{\rm search} = \theta_{INTEGRAL}$ = $3.6^{\prime}$.  If the source is outside the INTEGRAL error circle, then $\theta_{\rm search}$ is equal to the angular distance from the best estimate of the INTEGRAL position.  Another factor that is important in determining the probability that the Chandra/INTEGRAL association is spurious is the brightness of the source.  

\begin{figure}
\plotone{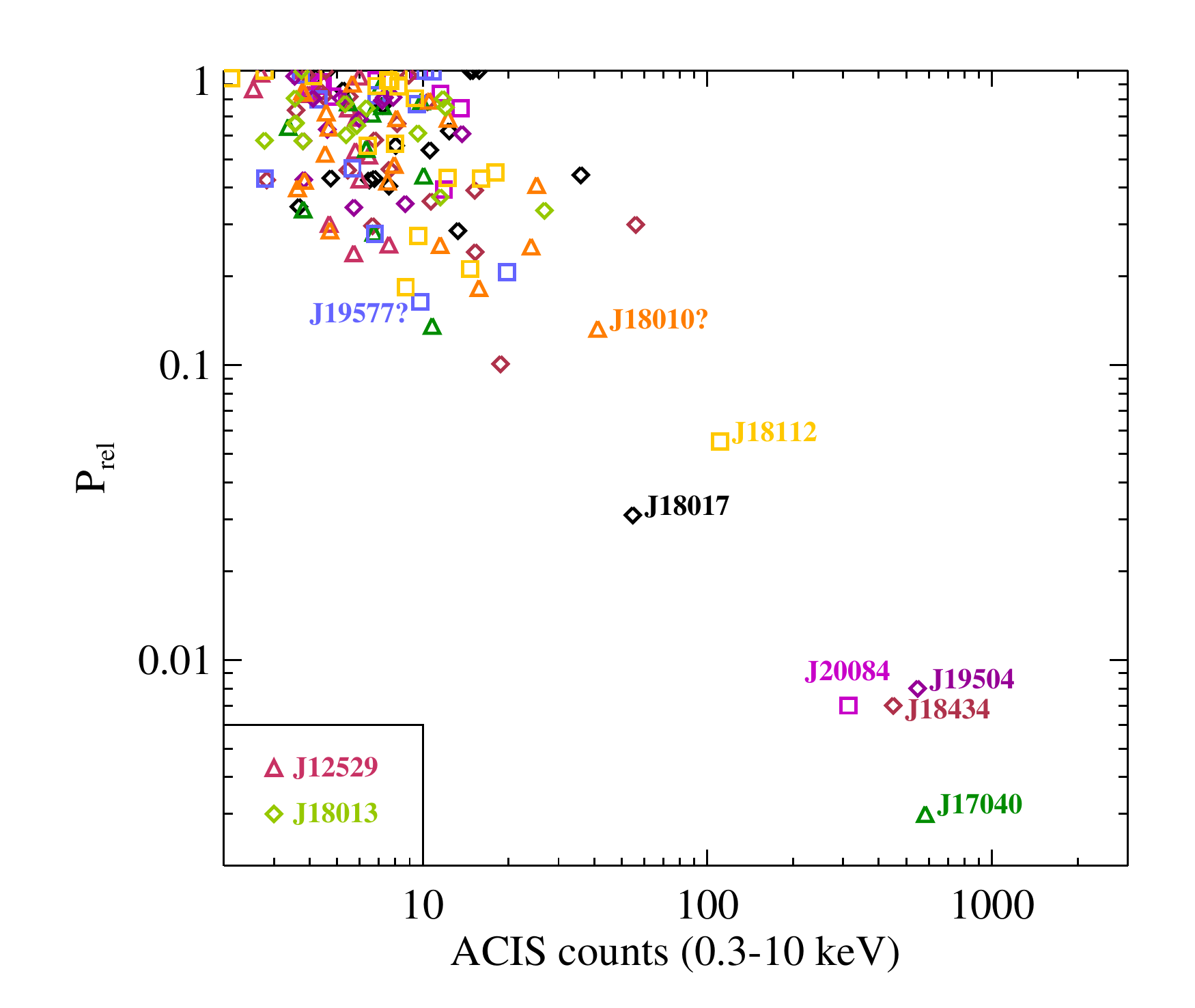}
\caption{The relative probability of a chance detection (see Equation 1) vs. the number of 0.3--10\,keV ACIS-I counts for all Chandra sources detected in the 10 observations of IGR source fields.  For the eight fields with candidate counterparts (although we note that J18010 and J19577 are questionable), the source that is least likely to be a spurious association is labeled with the IGR source name.  The legend in the lower left corner of the plot lists the two IGR source fields without likely counterparts.
\label{fig:prel}}
\end{figure}

As in \cite{tomsick20}, we determine the relative probabilities for all sources using
\begin{equation}
P_{\rm rel} = 1 - e^{-(\frac{C_{2-10\,\rm keV}}{C_{0}})^{-1.0}~\pi~\theta_{\rm search}^{2}}~~~,
\end{equation}
where $C_{2-10\,\rm keV}$ is the number of counts in the 2--10\,keV band, $C_{0}$ is a normalization constant set to a value of 140 so that the brightest sources have $P_{\rm rel}$ values near 1\%, and we use --1.0 as the slope of the $\log{N}$-$\log{S}$, which is intermediate between previously published profiles \citep{sugizaki01,fornasini14}.  In our case, with all of the observations having the same exposure, it is valid to use counts (as opposed to count rates) for this calculation.  The $P_{\rm rel}$ values for all the sources detected in the ten Chandra observations are plotted in Figure~\ref{fig:prel}.  This results in the field sources clustering at low numbers of counts and high values of $P_{\rm rel}$.  For six of the IGR fields (IGR~J17040--4305, IGR~J18017--3542, IGR~J18112--2641, IGR~J18434--0508, IGR~J19504+3318, and IGR~J20084+3221), there are clear Chandra counterparts that are well-separated from the field sources. There are also potential counterparts in the IGR~J18010--3045 and the IGR~J19577+3339 fields.  The IGR~J12529--6351 and IGR~J18013--3222 fields do not appear to have likely counterparts.

The candidate Chandra source in the IGR~J19577+3339 field has $P_{\rm rel} = 16.4$\%, which is the highest spurious association probability of any of the eight candidates we consider, and Figure~\ref{fig:prel} shows that there are even some field sources with lower probabilities.  However, searching in VizieR at this position finds a likely match with the radio source ICRF~J195740.5+333827.  The Chandra and radio Very Long Baseline Interferometry (VLBI) positions are consistent with a separation of $0.56^{\prime\prime}\pm 0.93^{\prime\prime}$ (90\% confidence). This is a bright radio source with a flux of $295\pm 9$\,mJy at 1.4\,GHz that appears in a number of radio catalogs.  The radio source is suspected to be an AGN, but no redshift has been measured. Although the X-ray source is faint, the combination of the hardness of its spectrum, the fact that it is only $1.12^{\prime}$ from the center of the INTEGRAL error circle, and the likely association with the bright radio source make it necessary for us to consider this candidate.

For the eight candidates, Table~\ref{tab:positions} gives the Chandra names and positions, the angular distance from the center of the INTEGRAL error circle ($\theta$), the number of ACIS counts in the 0.3--10\,keV band, and the hardness ratio. The uncertainties in the candidate positions include systematic and statistical contributions added in quadrature. We use the standard $0.8''$ systematic uncertainty (at the 90\% confidence level)\footnote{See http://cxc.cfa.harvard.edu/cal/ASPECT/celmon/} for each source. For the statistical uncertainty, we calculate the 90\% confidence intervals following equation 13 from \cite{kim07a}, which uses the number of counts as well as the angular distance of the source from the Chandra aimpoint. The Chandra counterpart to IGR~J18112--2641 has the greatest separation from its INTEGRAL position, at $6.05^{\prime}$. We also include the potential counterpart to IGR~J18010--3045 in Table~\ref{tab:positions} as well as in the rest of our analysis, although we cannot be as confident (with fewer counts and a separation of $5.95^{\prime}$) that the detected Chandra source is a true match.

In Section 3, we determine fluxes for the eight most likely counterpart candidates and use previously measured surface density ($\log{N}$-$\log{S}$) profiles \citep{sugizaki01,fornasini14} to calculate absolute spurious probabilities for the Chandra/INTEGRAL associations.

\section{Chandra Energy Spectra}

For the eight candidate counterparts, we extracted Chandra energy spectra using {\ttfamily specextract}.  We used the same source regions (circles with radii corresponding to 95\% encircled energy) and background regions as for the photometry.  We rebinned the spectra with the requirement of a detection in each bin at the 3--5$\sigma$ level, depending on the total number of counts in the spectrum.  The one exception is J19577, for which we binned to 1.5$\sigma$ because the spectrum only includes 10 counts.  We used XSPEC \citep{arnaud96} to fit the Chandra spectra with an absorbed power-law model, and the parameters are reported in Table~\ref{tab:spectra_ch}.  One of the bright sources, J17040, was also located close to the center of the field of view where the PSF is small.  This resulted in significant photon pile-up, and we included the XSPEC model {\ttfamily pileup} \citep{davis01} to account for this in the spectral fitting.  Photon pile-up does not impact the spectra of the other sources.  We performed the fits by minimizing the C-statistic, and we give the $C$ values and number of degrees of freedom (dof) in Table~\ref{tab:spectra_ch}.  

We use the $C$ values and the variances in $C$ \citep[see][]{kaastra17} to assess the quality of the fits.  The $P_{\rm reject}$ values in Table~\ref{tab:spectra_ch} indicate the probability that an absorbed power-law does not provide a good description of the spectrum.  The $P_{\rm reject}$ value is only meaningful if $C$ is larger than the number of dof, and this is only the case for J17040 (41\%), J19504 (41\%), and J20084 (77\%).  For these three sources the largest residuals are single bins at 6.7\,keV (3.6$\sigma$ above the continuum), 1.7\,keV (2.8$\sigma$ above the continuum), and 6.5\,keV (2.7$\sigma$ above the continuum), respectively.  While there is no immediate interpretation for the 1.7\,keV residual, it is possible that J17040 and J20084 have iron K$\alpha$ emission lines.  

Adding a narrow line for J17040 improves the fit to $C$/dof = 11/17, and the parameters, with 90\% confidence uncertainties, are $E_{\rm line} = 6.7\pm 0.1$\,keV and $N_{\rm line} = (6.4^{+19}_{-3.6})\times 10^{-5}$\,ph\,cm$^{-2}$\,s$^{-1}$.  The equivalent width is $EW = 1.5^{+8.5}_{-1.5}$\,keV, where the uncertainties come from simulations using the {\ttfamily eqwidth} command in XSPEC.  Although the improvement in the fit and the fact that $N_{\rm line}$ is greater than zero at the 90\% confidence level indicate that an iron line may be present, this is not a robust detection for at least three reasons: 1. The small value of $C$ relative to the number of dof suggests that adding the emission line may be over-fitting the spectrum; 2. $P_{\rm reject} = 41$\% indicates that it is fairly likely (59\%) that a power-law provides a good description of the spectrum; and 3. The simulations for determining the $EW$ indicate that the data are consistent with $EW=0$ at 90\% confidence.  

For J20084, adding a narrow line improves the fit to $C$/dof = 7/7, and the parameters are $E_{\rm line} = 6.7^{+0.1}_{-0.5}$\,keV, $N_{\rm line} = (3.1^{+2.9}_{-2.4})\times 10^{-5}$\,ph\,cm$^{-2}$\,s$^{-1}$, and $EW = 0.9^{+1.3}_{-0.8}$\,keV.  We estimate the significance of the emission line by determining the confidence level that makes the $N_{\rm line}$ error range consistent with zero, which is $\Delta C = 7.7$.  This indicates a significance of approximately 99.2\%, which corresponds to a 2.6$\sigma$ detection.  

We use the fluxes resulting from the spectral fits to determine the absolute spurious source probabilities for the eight Chandra candidates.  As in \cite{tomsick20}, we use
\begin{equation}
N(>F_{2-10~\rm keV,abs}) = 9.2(F_{2-10~\rm keV,abs}/10^{-13})^{-0.79}\,{\rm deg}^{-2}~~~,
\end{equation}
which is based on the $\log{N}$-$\log{S}$ curve for Galactic sources detected by the Advanced Satellite for Cosmology and Astrophysics \citep[ASCA;][]{sugizaki01}, and we also use 
\begin{equation}
N(>F_{2-10~\rm keV,unabs}) = 36(F_{2-10~\rm keV,unabs}/10^{-13})^{-1.24}\,{\rm deg}^{-2}~~~,
\end{equation}
which is based on Chandra observations of the Norma region of the Galactic plane \citep{fornasini14} and includes Galactic sources and AGN.  In both cases, the 2--10\,keV flux range is used, and while \cite{sugizaki01} simply use the absorbed fluxes, \cite{fornasini14} make a correction for Galactic absorption.  Thus, from the power-law fits, we provide in Table~\ref{tab:spectra_ch} both the absorbed 2--10\,keV fluxes as well as fluxes with an absorption correction.  We correct for the best fit value of $N_{\rm H}$ unless it is larger than the Galactic value from \cite{bb16}, in which case, we correct for $N_{\rm H,Galactic}$ (also given in Table~\ref{tab:spectra_ch}). The resulting absorption corrections are relatively small.

With these fluxes and the two $\log{N}$-$\log{S}$ curves, we use
\begin{equation}
P = 1 - e^{-N(>F_{2-10~\rm keV})~\pi~\theta_{\rm search}^{2}}
\end{equation}
to calculate the absolute probability of finding a source in the search region (defined by $\theta_{\rm search}$ as described in Section 2.1) as bright as the candidate sources by chance.  The ranges of probabilities (from the two $\log{N}$-$\log{S}$ curves) for each source are given in Table~\ref{tab:spectra_ch}.  This shows that the candidate Chandra counterparts to J17040, J20084, and J18454 are very unlikely ($<$1\%) to be detected in the search regions by chance.  Also, the candidates to J19504 and J18017 are unlikely (one to a few percent) to be chance detections.  The J18112, J19577, and J18010 candidates have somewhat higher spurious probabilities (5.8--8.9\%, 4.9--12\%, and 10--21\%, respectively), and although, in the following, we perform the same analyses for these three as for the other five, they may possibly be field sources rather than the actual counterpart to the IGR sources (although see Section 2.1 concerning the association of J19577 with a bright radio source).

\begin{figure}
\plotone{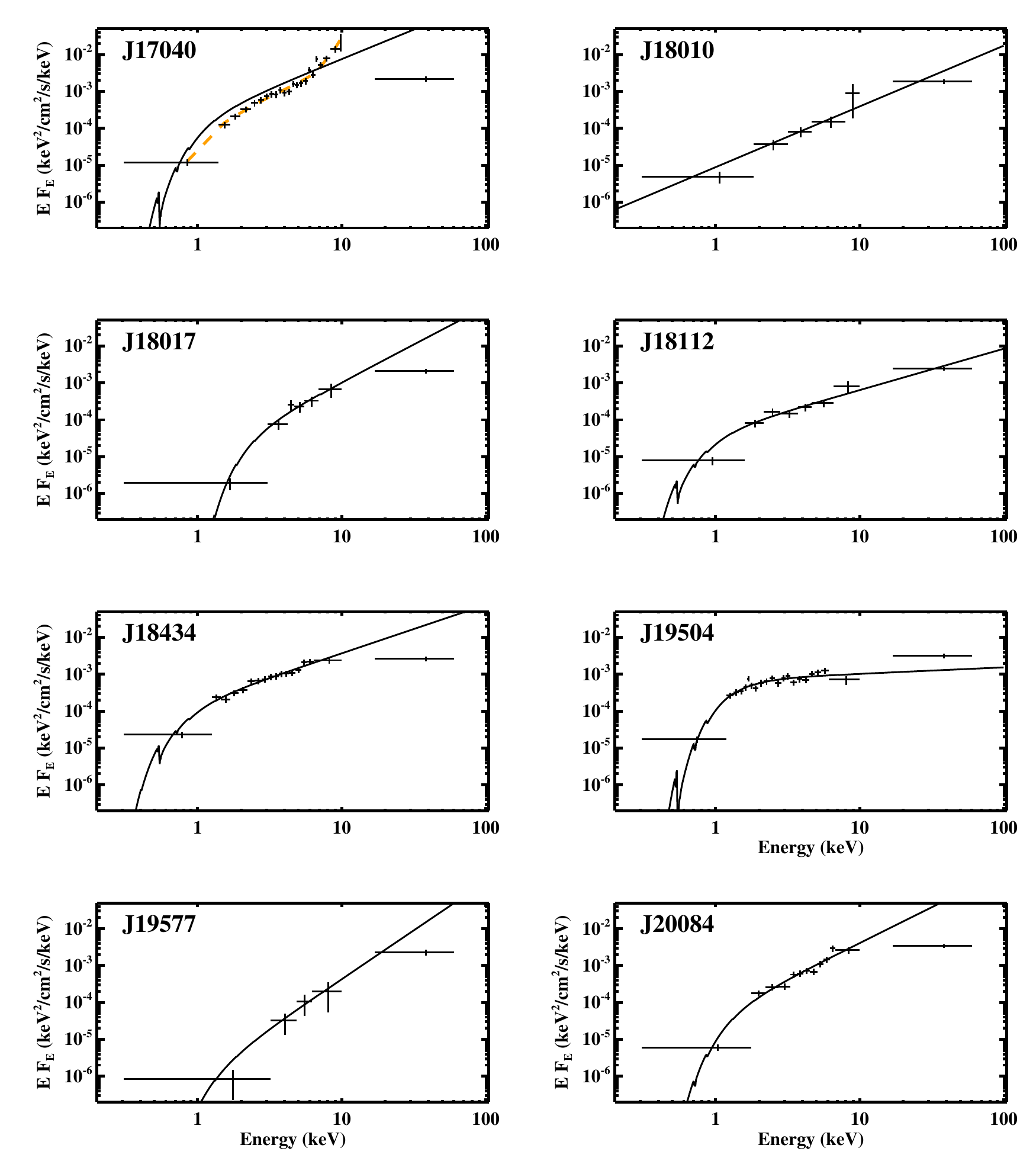}
\vspace{1.0cm}
\caption{Chandra energy spectra for the eight candidate counterparts along with the 17--60\,keV INTEGRAL flux point from \cite{krivonos17}.  The model shown is the best fit absorbed power-law to the Chandra data.  The power-law is extended to higher energies to compare the extrapolation to the INTEGRAL flux point.  For J17040, the dashed orange curve shows the fit to the Chandra spectrum before the pile-up correction.
\label{fig:spectra}}
\end{figure}

Figure~\ref{fig:spectra} shows the Chandra spectra with the absorbed power-law fits.  As given in Table~\ref{tab:spectra_ch}, the spectra have hard power-law photon indices with best fit values between $\Gamma$ = --0.6 and 1.8.  In Figure~\ref{fig:spectra}, we have added a 17--60\,keV point measured by INTEGRAL \citep{krivonos17} for comparison to the extrapolation of the power-law. In making this comparison, it is important to keep in mind that the INTEGRAL flux point is an average over 14 years while the Chandra spectra are a single observation.  Thus, differences can be related to source variability or to a change in spectral slope between the Chandra band and the INTEGRAL band.  The 17--60\,keV flux point is lower than the extrapolation of the power-law for J17040, J18010, J18017, J18434, J19577, and J20084, which makes it possible that the spectrum has a break or a cutoff above 10\,keV.  These are also the six sources with the hardest power-law indices (best fit $\Gamma$ values between --0.6 and 0.7).  For J18112, the 17--60\,keV point is consistent with the power-law extrapolation.  

Although the softest source (J19504 with $\Gamma = 1.8\pm 0.3$) has a 17--60\,keV point that is higher than the power-law extrapolation, a re-inspection of the INTEGRAL image shows significant noise in the part of the sky where J19504 lies due to the proximity to the bright source Cygnus~X-1.  In fact, J19504 clearly sits on top of a positive noise artifact, indicating that the INTEGRAL flux reported in \cite{krivonos17} and shown in Figure~\ref{fig:spectra} is an overestimation.

\section{Optical/IR Identifications}

We used the VizieR database to search for optical/IR counterparts to the eight Chandra sources. In Table~\ref{tab:gaia3}, we provide details on the matches found using Gaia EDR3, including the $G$-, $BP$-, and $RP$-band magnitudes, the parallax, the astrometric noise, and the proper motion. For three of the sources, a reliable distance measurement is available, and we quote the geometric distances\footnote{Although the \cite{bj21} catalog includes both geometric and photogeometric distances, we use the geometric distances because the colors of the sources we are studying may deviate from the assumptions made for the photogeometric distances.} calculated by \cite{bj21}. Those distances are $0.937\pm 0.046$\,kpc, $1.58\pm 0.13$\,kpc, and $3.0^{+1.2}_{-0.9}$\,kpc, for J17040, J18017, and J18434, respectively.  These three sources also have high proper motions as do J18112 ($5.74\pm 0.09$ milliarcsec/yr) and J20084 ($4.77\pm 0.23$ milliarcsec/yr).  Thus, we conclude that all five of these sources are Galactic. J19504 has a proper motion of $0.12\pm 0.17$ milliarcsec/year, which is consistent with zero.  The small proper motion does not distinguish between a Galactic or extragalactic nature for J19504, but it does allow for the possibility that the source is extragalactic.  The Gaia counterpart to J18010 does not have parallax or proper motion measurements, presumably because it is too faint.  J19577 does not have a Gaia counterpart.  The closest Gaia source is $1.5^{\prime\prime}$ away.

Although the \cite{bj21} catalog provides distance estimates for J18112, J19504, and J20084, it also provides information about these estimates suggesting that they may not be reliable.  For J18112 and J19504, the astrometric noise is larger than the value of the parallax (Table~\ref{tab:gaia3}).  For J20084, while there is no astrometric noise, the uncertainty on the parallax is almost twice that of the parallax value, and Figure~6 in \cite{bj21} shows that, in such cases, the distance estimate is highly dependent on the prior distribution assumed.

We also used VizieR to search for AllWISE \citep{cutri21} IR counterparts to the eight Chandra sources. We found just two matches in the AllWISE catalog.  These are for the Chandra counterparts to J19504 and J20084. Both AllWISE counterparts are consistent with being point-sources according to the extended source parameter "ex," which has a value of zero in both cases.  The J19504 counterpart is AllWISE~J195019.73+331416.3, which has magnitudes of $W1 = 12.057\pm 0.025$, $W2 = 11.209\pm 0.020$, $W3 = 9.028\pm 0.026$, and $W4 = 7.14\pm 0.10$.  The $W1$-$W2$ and $W2$-$W3$ colors place J19504 is a region of the near-IR color plot that is commonly populated by AGN \citep{mateos12,secrest15}.  We have also checked the near-IR colors for J20084, but it does not fall in the AGN region.  The positions of both of the AllWISE sources are consistent with the Gaia positions to within $<$0.3$^{\prime\prime}$.

\begin{figure}
\epsscale{1.2}
\hspace{-2cm}
\plotone{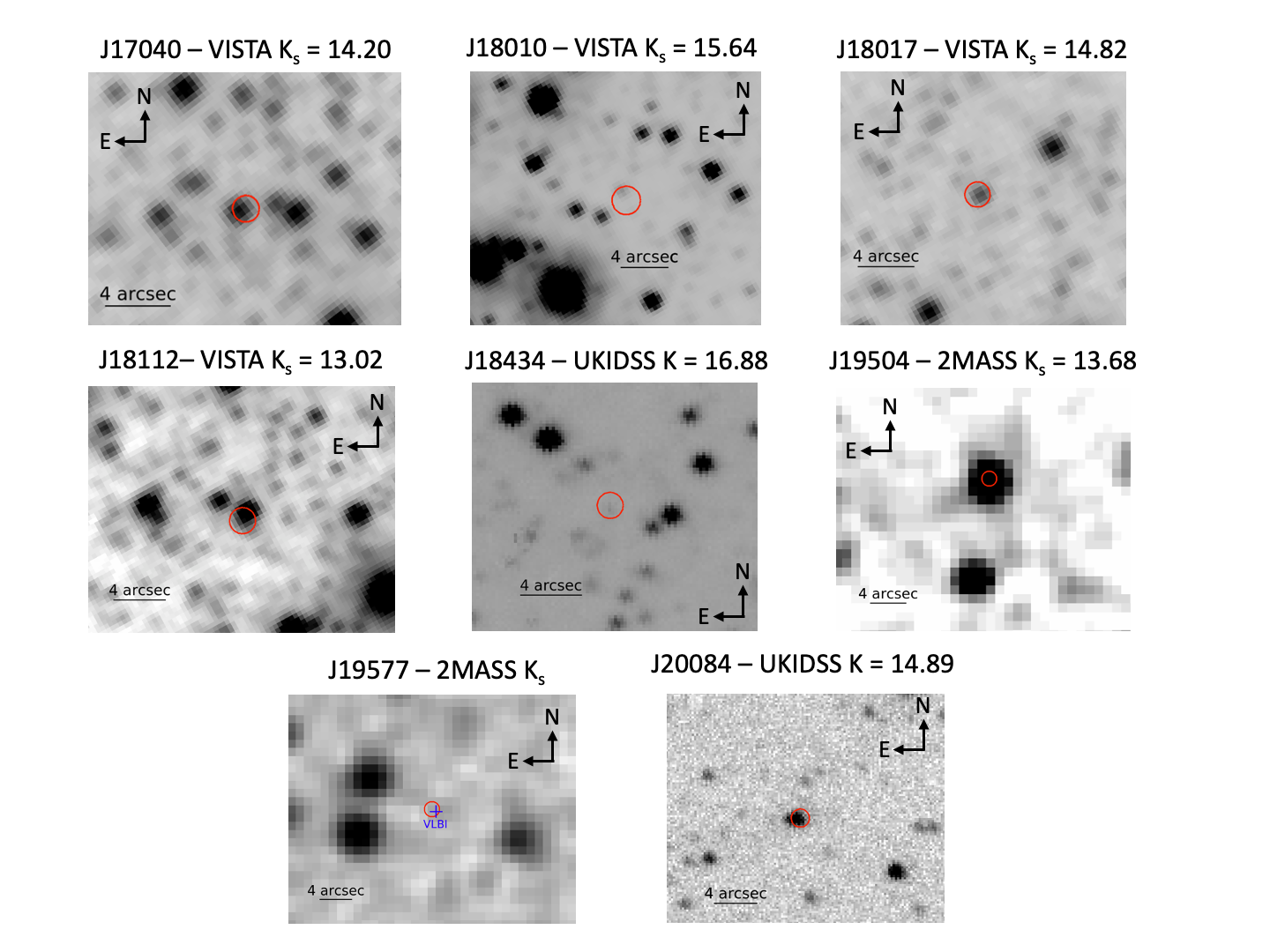}
\caption{$K$ or $K_{s}$ band images from the VISTA VVV, 2MASS, and UKIDSS surveys for the eight IGR sources with candidate Chandra counterparts.  The red circles indicate the Chandra positions (90\% confidence).  For J19577, there is no 2MASS source consistent with the Chandra position, but there is a VLBI radio source (marked with a blue cross) in the Chandra error circle. Although J19504 appears to be a single source at 2MASS resolution, the PanSTARRS image in Figure~\ref{fig:nir_y} shows that it is blended. \label{fig:nir_k}}
\end{figure}

\begin{figure}
\epsscale{1.2}
\hspace{-2cm}
\plotone{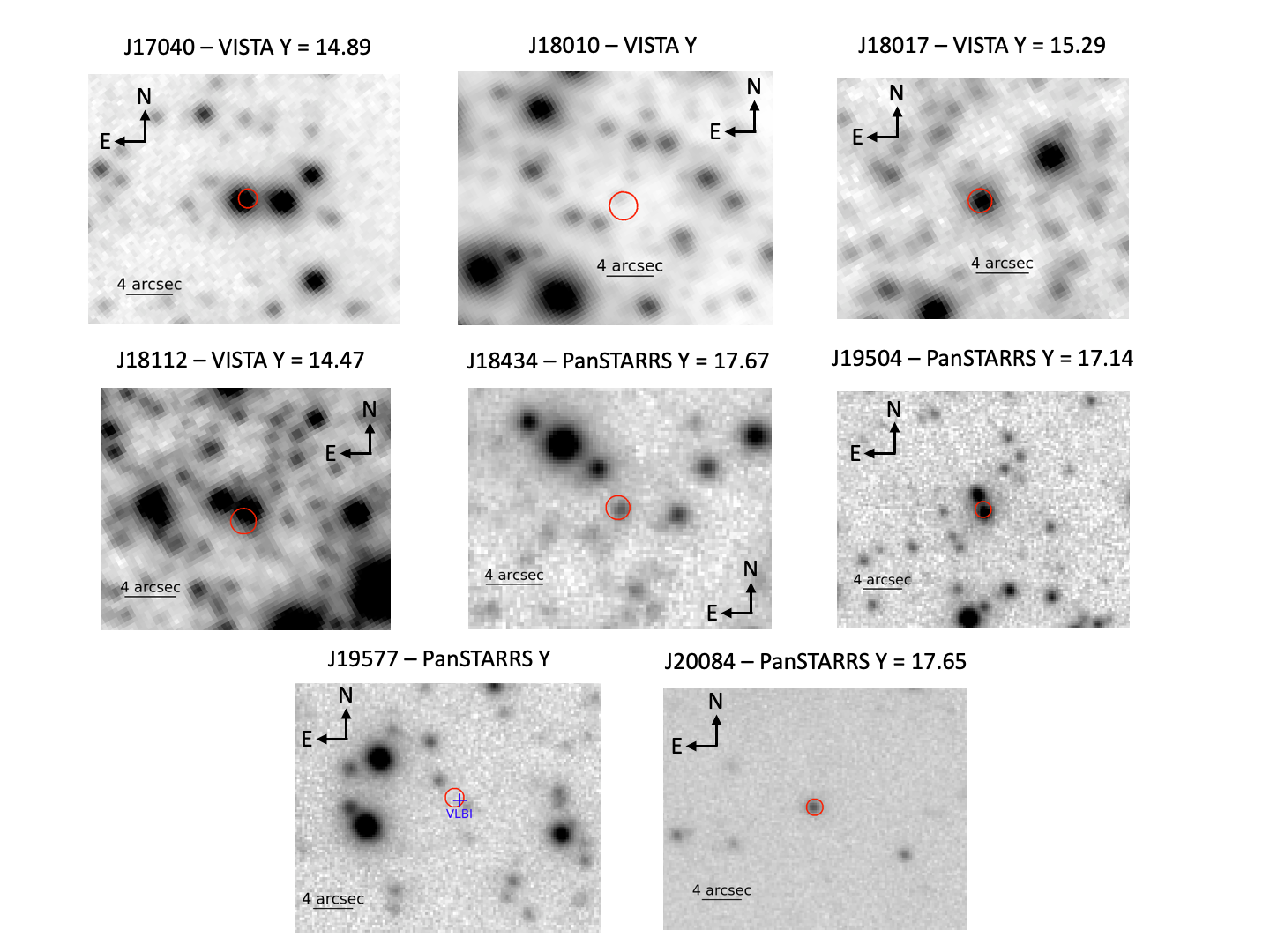}
\caption{$Y$ band images from the VISTA VVV and PanSTARRS surveys for the eight IGR sources with candidate Chandra counterparts.  The red circles indicate the Chandra positions (90\% confidence).  For J18010, although the near-IR counterpart is marginally visible, no $Y$-band magnitude is given in the VISTA VVV DR2 data base.
\label{fig:nir_y}}
\end{figure}

In Table~\ref{tab:nir}, we include the other identifications from the results of VizieR searches, focusing our results to the near-IR matches. These include those matches found in VISTA VVV \citep{minniti10, mcmahon13, minniti17}, the 2 Micron All-Sky Survey, \citep[2MASS;][]{cutri03}, UKIDSS \citep{lucas08}, and PanSTARRS \citep{chambers16}. Across these four survey catalogs, we have measurements of the source magnitudes in the $Y$, $J$, $H$, and $K$/$K_{s}$ bands, except for J18010, which does not have a reliable $Y$-band measurement due to the source being too faint at those wavelengths. In all cases, the near-IR positions are consistent with the Gaia positions.  In Figure~\ref{fig:nir_k} we provide the $K$/$K_{s}$ images of the region in the sky covering each Chandra counterpart’s position, marked with a red circle in each image. Figure~\ref{fig:nir_y} provides the $Y$-band images for each of the eight Chandra counterparts as well.  The images for J19577 show that there is no optical or near-IR counterpart in the Chandra error circle.  Also, the VLBI radio position, which is inside the Chandra error circle, is marked.

In summary, by utilizing the Chandra counterpart positions and searching the VizieR database, we are able to find optical/IR counterparts for seven sources and a radio counterpart for J19577. From the Gaia EDR3 data, we established distances to three sources, while the proper motions of two of the remaining sources indicate that they are also Galactic in nature.  For J19504, the fact that the AllWISE colors place it in a region populated by AGN suggest that J19504 is an AGN, and while the low proper motion in Gaia does not prove that the source is extragalactic, it is consistent with that interpretation. The near-IR counterpart information, including magnitudes, is provided in Table~\ref{tab:nir} and Figures~\ref{fig:nir_k} and \ref{fig:nir_y}.

The VizieR search also uncovered counterparts in the VST\footnote{Very Large Telescope (VLT) Survey Telescope} Photometric H$\alpha$ Survey of the Southern Galactic Plane and Bulge (VPHAS+ Data Release 2) for J17040 and J18434 \citep{drew16}.  For J17040, the counterpart is VPHASDR2 J170405.0--430538.0 with magnitudes of $15.68\pm 0.01$ for the H$\alpha$ filter, $r = 16.10\pm 0.01$, and $i = 15.47\pm 0.01$.  For J18434, the counterpart is VPHASDR2 J184311.4--050545.5 with magnitudes of $17.83\pm 0.02$ (H$\alpha$), $r = 18.31\pm 0.01$, and $i = 17.71\pm 0.01$.  These correspond to $r$--H$\alpha$ = $0.42\pm 0.01$ and $r$--$i$ = $0.63\pm 0.01$ for J17040 and $r$--H$\alpha$ = $0.48\pm 0.02$ and $r$--$i$ = $0.60\pm 0.01$ for J18434.  Comparing to the field star distribution in figure 17 of \cite{drew14}, these measurements indicate that J17040 and J18434 have an excess at H$\alpha$, suggesting the presence of an emission line.

\section{Sources without clear Chandra counterparts to the IGR sources}

The Chandra observations for IGR~J18013--3222 and IGR~J12529--6351 did not lead to detections of clear counterparts, but here we consider the Chandra sources in each field with the lowest values of $P_{\rm rel}$.

For J18013, the sources with the lowest values of $P_{\rm rel}$ have values of 37.2\% and 33.5\%, indicating that there is a high chance that they are spurious.  In addition, the one with $P_{\rm rel} = 33.5$\%, CXOU~J180143.1--321540, is $8.4^{\prime}$ from the center of the INTEGRAL error circle, which is another reason to doubt that it is the correct counterpart. Within the $3.6^{\prime}$ INTEGRAL error circle, the brightest Chandra source only has 4 counts.  We conclude that the upper limit on the 2--10\,keV flux is $<$$1\times 10^{-13}$\,erg\,cm$^{-2}$\,s$^{-1}$. 

For J12529, CXOU~J125231.0--635021 is within the INTEGRAL error circle and has $P_{\rm rel} = 24$\% and $5.7\pm 3.6$ detected ACIS counts.  While this is a small number, they are all $>2$\,keV, suggesting that it is a hard source.  It is possible that this is the correct counterpart of the INTEGRAL source, but the evidence is not strong enough to consider it as a likely counterpart.  We conclude that the upper limit on the 2--10\,keV flux is $<$$1\times 10^{-13}$\,erg\,cm$^{-2}$\,s$^{-1}$. 

\section{Discussion}

From Chandra observations of 10 IGR sources, we have found definite or candidate soft X-ray counterparts in eight cases.  The Chandra positions provide information about the multi-wavelength properties of these sources, and we discuss the nature of the sources based on the X-ray and multi-wavelength information.

\subsection{Galactic Sources with Distances}

J17040, J18017, and J18434 are the three sources with distance constraints from Gaia (Table~\ref{tab:gaia3}).  Based on the X-ray fluxes measured by Chandra (Table~\ref{tab:spectra_ch}) the 2--10\,keV luminosities are $(7.9\pm 0.8)\times 10^{32}$, $(2.1\pm 0.4)\times 10^{32}$, and ($4.2^{+3.4}_{-2.6})\times 10^{33}$\,erg\,s$^{-1}$, respectively.  Based on these X-ray luminosities and the hardness of the spectra, the emission is not from isolated stars (consistent with them being detected by INTEGRAL), and the most likely possibility would be that these are accreting binaries, such as CVs or X-ray binaries.  For sources with X-ray spectra with power-law photon indices of $\Gamma < 1$, there are two types of binaries that are the most likely possibilities:  magnetic CVs, such as intermediate polars, and HMXBs with highly magnetized neutron stars.  The magnetic CVs typically have late-type donor stars, while the HMXBs have O or B type stars (by definition).  In the following, we use the distances and near-IR magnitudes to consider the most likely classifications for J17040, J18017, and J18434.

Interpreting the near-IR magnitudes of these sources (Table~\ref{tab:nir}) requires that we correct them for Galactic extinction, and we have used 3D dust maps to determine this.  For J17040 and J18017, we use the {\ttfamily mwdust} code\footnote{See https://github.com/jobovy/mwdust} \citep{bovy16} to obtain the $E(B-V)$ values shown in Table~\ref{tab:distances}.  The errors come from the $E(B-V)$ range for the Gaia distance range.  For J18434, we use the Bayestar19 map from \cite{green19}, accessing $E(g-r)$ values using a web interface\footnote{See http://argonaut.skymaps.info/query} and multiplying by 0.94 to convert to $E(B-V)$\footnote{See http://argonaut.skymaps.info/usage}.  In Table~\ref{tab:distances}, $A_{V} = 3.1 E(B-V)$, $A_{J} = 0.260 A_{V}$, and $A_{J}$--$A_{H}$ = 0.090 $A_{V}$ \citep{fitzpatrick99}.  We correct the $J$ and $H$ magnitudes in Table~\ref{tab:nir} for extinction, obtaining $J$--$H$--($A_{J}$-$A_{H}$) values of $0.061^{+0.007}_{-0.008}$, --$0.030^{+0.016}_{-0.017}$, and $0.63\pm 0.27$ for J17040, J18017, and J18434, respectively.  

Using tables of stellar colors from \cite{pm13}, we find that the measured colors correspond to temperatures of $T_{\rm eff} = 7680\pm 90$, $9600\pm 800$, and 2200--5300\,K for the three sources (Table~\ref{tab:nir}).  We also calculated the absolute $J$-band magnitudes, $M_{J}$, using the values of $J$, $A_{J}$, and the Gaia distances.  Although we do not know if the near-IR emission is coming from a star or not, the values of $M_{J}$ place limits on the main sequence spectral types of a star in the system, and we determine these limits using tables of main sequence stellar magnitudes and colors \citep{pm13}.  As shown in Table~\ref{tab:distances}, the spectral type limits for J17040 and J18017 indicate temperatures of $<$5100 and $<$5770\,K, respectively, which are less than the values derived from the colors.  This suggests that the near-IR is dominated by hot emission from an accretion disk.  On the other hand, for J18434, $M_{J}$ indicates a temperature limit of $<$4400\,K, which is consistent with the color temperature.  Thus, based on this analysis, it is possible that the near-IR emission for J18434 comes from a cool star.

We repeated the color and magnitude analysis using the optical magnitudes from Gaia. In Table~\ref{tab:distances}, $A_{G} = 0.90 A_{V}$ and $A_{BP}$--$A_{RP}$ = 0.51 $A_{V}$ \citep{fitzpatrick99}.  For J17040 and J18017, the results are similar to the near-IR in that they show higher values of $T_{\rm eff}$ based on the colors than the temperature ($T_{\rm type}$) of a star of spectral type consistent with the observed absolute magnitude.  Although the $T_{\rm eff}$ temperatures from the optical are not formally consistent with the near-IR, the measurements are still consistent with emission from an accretion disk, and the different temperatures may be caused by source variability.  In addition, for J18434, $T_{\rm eff} > 7900$\,K while $T_{\rm type}$ is constrained to be lower.  Since the Gaia and near-IR measurements are made at different times, the likely interpretation is that J18434 also has emission from a variable accretion disk.  A comparison of the VISTA and PanSTARRS magnitudes provides additional evidence that J18434 is variable in the optical and near-IR (Table~\ref{tab:nir}).  The evidence for H$\alpha$ emission lines for J17040 and J18434 from the VPHAS+ survey presented in Section~4 is consistent with emission from an accretion disk for these two sources.

Given the presence of hard X-ray emission, finding evidence for an accretion disk is expected.  The presence of accretion, the evidence for late-type companions, and the very hard values of the X-ray power-law index suggests a possible magnetic CV nature.  For J17040, $\Gamma = 0.2\pm 0.4$, which indicates that a magnetic CV nature is very likely.  The power-law photon index is not as well constrained for J18017 ($\Gamma = 0.0^{+1.7}_{-1.3}$), but the source may also be a magnetic CV.  At $\Gamma = 0.7\pm 0.3$, J18434 is somewhat softer than J17040, but we still consider it to be a strong magnetic CV candidate.

\subsection{Other Galactic Sources}

J18112 and J20084 both have large Gaia-measured proper motions and are Galactic, but the source distances are not known.   Both have hard spectra with $\Gamma = 0.9^{+0.7}_{-0.5}$ for J18112 and $0.1^{+0.4}_{-0.2}$ for J20084, suggesting that they may be magnetic CVs or HMXBs.

In the direction of J18112 ($l = 4.78^{\circ}$, $b$ = --3.83$^{\circ}$, which is in the Galactic bulge), $E(B-V)$ increases from 0.14 at a distance ($d$) of 500\,pc and reaches a maximum of 0.68 at $d = 4$\,kpc \citep{bovy16}.  At 500\,pc and using the near-IR magnitudes in Table~\ref{tab:nir}, $J$--$H$--($A_{J}$--$A_{H}$) = 0.615, corresponding to $T_{\rm eff} = 3660$\,K, which is the temperature for an early M-type star.  The absolute magnitude at 500\,pc would be $M_{J} = 5.34$, corresponding to a late K-type main sequence star. It is clear that the nearby distance scenario strongly disfavors the HMXB possibility.  For distances $>$4\,kpc, $J$--$H$--($A_{J}$--$A_{H}$) = $0.463\pm 0.011$, corresponding to $T_{\rm eff}$ = 4830--5100\,K (K2V-K3V).  The absolute magnitude is $M_{J} < 0.39$ for $d > 4$\,kpc.  While the absolute magnitude at large distances allows for the presence of a high-mass companion star, the low temperature derived from the colors is inconsistent with this scenario.  Thus, we conclude that a magnetic CV with a low-mass star is strongly favored.  The brightness in the near-IR may indicate that the companion is an evolved giant star.  However, there also may be a component of the emission from an accretion disk based on the fact that J18112 is variable in the $Y$-band (Table~\ref{tab:nir}).

J20084 is essentially in the middle of the Galactic plane ($b$ = --0.23$^{\circ}$), and it has relatively high extinction.  Using \cite{green19}, we find $E(B-V) > 1.5$ if the distance is $>$1.8\,kpc and $E(B-V) > 2.0$ if $d > 4.5$\,kpc.  Given the near-IR magnitudes (Table~\ref{tab:nir}), $J$--$H$--($A_{J}$--$A_{H}$) is $<$0.30 at the lower distance and $<$0.16 for larger distances.  These colors correspond to temperature limits of $T_{\rm eff} > 5900$\,K and $>$6700\,K.  Without a distance, $M_{J}$ is highly uncertain.  At a distance of 4.5\,kpc and $A_{J} = 1.6$, $M_{J} = 1.2$, which would indicate an early-A spectral type if the stellar component dominates the near-IR emission.  However, the distance could be larger, which would make the spectral type even earlier.  Thus, we cannot rule out the possibility that J20084 is an HMXB.

\subsection{Sources with Likely AGN Classifications}

The evidence that J19504 is an AGN is primarily based on the fact that its WISE colors place it in the AGN region.  Although it is marked as an extended source in PanSTARRS (Table~\ref{tab:nir}), it is unclear whether it is truly extended or if the nearby and partially blended star (Figure~\ref{fig:nir_y}) is the reason it is flagged as extended in the PanSTARRS processing.  The Gaia measurements give a negative parallax and a proper motion consistent with zero (Table~\ref{tab:gaia3}), which is consistent with but does not prove the AGN hypothesis. J19504 is also an outlier in it X-ray spectral parameters with a softer power-law slope ($\Gamma = 1.8\pm 0.3$) and possible evidence for absorption greater than the Galactic value (Table~\ref{tab:spectra_ch}).  We conclude that IGR~J19504+3318 is very likely to be an AGN, but more proof is needed.

For J19577, even though the flux of the Chandra counterpart allows for the possibility that it is unrelated to the INTEGRAL source, the fact that the source is actually more unusual in being a 295\,mJy radio source increases the likelihood that it is related to IGR~J19577+3339.  The radio source does not have a redshift. We conclude that IGR~J19577+3339 is likely to be an AGN.

\subsection{Unclassified Sources}

While three sources remain unclassified (J18010, J18013, and J12529), this study has provided useful information.  For J18010, there is a candidate Chandra counterpart with a relatively hard spectrum that extrapolates to a flux in the INTEGRAL bandpass that is consistent with the flux measured by INTEGRAL (Figure~\ref{fig:spectra}).  In addition, the Chandra position is consistent with an optical/near-IR source detected by Gaia (although it is too faint for a distance or proper motion measurement) and VISTA.  The $K_{s}$-band magnitude is $15.640\pm 0.095$, which is bright enough to obtain a near-IR spectrum, which would greatly help in the classification of this source.  We also note that the source is $<$$4^{\circ}$ from the center of the Galaxy, which increases the probability that it is Galactic.  

Neither J18013 nor J12529 have likely Chandra counterparts.  The flux upper limits are low enough compared to the flux measured by INTEGRAL that the source is likely to be variable or transient.  However, since the Chandra and INTEGRAL bands do not overlap, we cannot rule out the possibility that Chandra does not detect the source due to a high column density.

\section{Summary and Conclusions}

The larger context for this work is the INTEGRAL Galactic plane surveys, which are providing new information about hard X-ray Galactic populations. Table~\ref{tab:summary} provides a summary of the results obtained using the 10 Chandra observations.  In five cases, the Chandra localization identifies a Gaia source with a distance or high proper motion measurement, indicating that the sources are Galactic. They all have hard X-ray spectra, suggesting that they are either magnetic CVs or HMXBs.  For the three sources with distance measurements (IGR~J17040--4305, IGR~J18017--3542, and IGR~J18434--0508), we argue that the magnetic CV classification is the most likely one.  For the other two Galactic sources, IGR~J18112--2641 is also more likely to be a CV than an HMXB based on analysis of the near-IR magnitudes and the fact that the source shows variability by a magnitude in the $Y$-band.  IGR~J20084+3221 could be a CV or an HMXB.  IGR~J19504+3318 and IGR~J19577+3339 are candidate AGN.  With the exception of J19577, the sources are bright enough for optical or near-IR spectroscopy, which will provide definitive classifications.

\acknowledgments

JAT and BC acknowledge partial support from the National Aeronautics and Space Administration (NASA) through Chandra Award Numbers GO7-18030X and GO8-19030X issued by the Chandra X-ray Observatory Center, which is operated by the Smithsonian Astrophysical Observatory under NASA contract NAS8-03060. JH acknowledges support from an appointment to the NASA Postdoctoral Program at the Goddard Space Flight Center, administered by the USRA through a contract with NASA.  SC is grateful to the Centre National d'Etudes Spatiales (CNES) for the funding of MINE (Multi-wavelength INTEGRAL Network). MC and JR acknowledge financial support from CNES.  RK acknowledges support from the Russian Science Foundation (grant 19-12-00396). This research has made use of the VizieR catalog access tool and the SIMBAD database, which are both operated at CDS, Strasbourg, France.  This work has made use of data from the European Space Agency (ESA) mission Gaia (\url{https://www.cosmos.esa.int/gaia}), processed by the Gaia Data Processing and Analysis Consortium (DPAC, \url{https://www.cosmos.esa.int/web/gaia/dpac/consortium}). Funding for the DPAC has been provided by national institutions, in particular the institutions participating in the Gaia Multilateral Agreement. This publication makes use of data products from the Wide-field Infrared Survey Explorer (WISE), which is a joint project of the University of California, Los Angeles, and the Jet Propulsion Laboratory/California Institute of Technology, funded by NASA.  The work has been partially based on data products from observations made with ESO Telescopes at the La Silla Paranal Observatory under program ID 179.B-2002 (VVV) and ID 177.D-3023 (VPHAS+, www.vphas.eu).  The work has also used results from the UKIDSS and PanSTARRS surveys.  The PanSTARRS Surveys were made possible through contributions by the Institute for Astronomy, the University of Hawaii, the PanSTARRS Project Office, the Max-Planck Society and its participating institutes, the Max Planck Institute for Astronomy, Heidelberg and the Max Planck Institute for Extraterrestrial Physics, Garching, The Johns Hopkins University, Durham University, the University of Edinburgh, the Queen's University Belfast, the Harvard-Smithsonian Center for Astrophysics, the Las Cumbres Observatory Global Telescope Network Incorporated, the National Central University of Taiwan, the Space Telescope Science Institute, and the NASA under Grant No. NNX08AR22G issued through the Planetary Science Division of the NASA Science Mission Directorate, the NSF Grant No. AST-1238877, the University of Maryland, Eotvos Lorand University (ELTE), and the Los Alamos National Laboratory.The Pan-STARRS1 Surveys are archived at the Space Telescope Science Institute (STScI) and can be accessed through MAST, the Mikulski Archive for Space Telescopes. Additional support for the PanSTARRS public science archive is provided by the Gordon and Betty Moore Foundation.

\smallskip
\facilities{CXO, INTEGRAL, Gaia, WISE, VLT, UKIRT}

\smallskip
\software{CIAO (Fruscione et al. 2006), XSPEC (Arnaud 1996)}




\begin{table}
\caption{Source information from the 2017 INTEGRAL catalog\label{tab:krivonos}}
\begin{minipage}{\linewidth}
\footnotesize
\begin{tabular}{ccccccc} \hline \hline
IGR Name & $l$\footnote{Galactic longitude converted from INTEGRAL position.}   & $b$\footnote{Galactic latitude converted from INTEGRAL position.}   &  RA\footnote{Source position measured by INTEGRAL and reported in \cite{krivonos17}.  Individual position uncertainties are not provided in \cite{krivonos07}, but it is indicated that the typical 90\% confidence INTEGRAL error radius for these sources is $3.6^{\prime}$.} & Dec$^{c}$ & Flux\footnote{The flux measured by INTEGRAL in units of $10^{-11}$\,erg\,cm$^{-2}$\,s$^{-1}$.}  & Significance\footnote{The significance of the INTEGRAL detection in terms of signal-to-noise.}\\
         & (deg) & (deg) &  (deg)   & (deg)    & (17--60\,keV) &             \\ \hline\hline
J20084+3221  & 70.04 & --0.23   & 302.124 & +32.350 & $0.68\pm 0.08$ & 8.4\\
J18434--0508 & 27.45 & --0.56   & 280.855 & --5.138 & $0.52\pm 0.08$ & 6.2\\
J12529--6351 & 303.10 & --1.00 & 193.241 & --63.868 & $0.49\pm 0.09$ & 5.5\\
J17040--4305 &  343.61 & --1.02 & 256.010 & --43.080 & $0.44\pm 0.07$ & 6.2\\
J19577+3339  &   69.95 &  +2.38 & 299.429 & +33.658 & $0.46\pm 0.08$ & 5.6\\
J19504+3318  &   68.87 &  +3.50 & 297.615 & +33.311 & $0.63\pm 0.09$ & 7.4\\
J18010--3045  &   0.12 & --3.82 & 270.271 & --30.764 & $0.37\pm 0.05$ & 7.2\\
J18112--2641  &   4.78 & --3.83 & 272.854 & --26.707 & $0.48\pm 0.06$ & 8.7\\
J18013--3222  & 358.74 & --4.65 & 270.326 & --32.371 & $0.34\pm 0.05$ & 6.4\\
J18017--3542 & 355.90 & --6.27 & 270.371 & --35.638 & $0.42\pm 0.06$ & 7.0\\ \hline
\end{tabular}
\end{minipage}
\end{table}

\begin{table}
\caption{{\em Chandra} Observations\label{tab:obs}}
\begin{minipage}{\linewidth}
\footnotesize
\begin{tabular}{cclccccc} \hline \hline
IGR Name & ObsID & Start Time (UT) & Exposure\\
         &       &                 & Time (s)\\ \hline\hline  
J18017--3542 & 20200 & 2018 Jul 15, 21.8\,h & 4956\\
J12529--6351 & 20201 & 2018 Feb 20, 3.3\,h & 4952\\
J20084+3221 & 21248 & 2019 Jan 9, 19.5\,h & 4949\\
J18434--0508 & 21249 & 2019 Mar 5, 3.4\,h & 4952\\
J17040--4305 & 21250 & 2019 May 29, 1.5\,h & 4956\\
J19577+3339  & 21251 & 2018 Nov 30, 8.7\,h & 4959\\
J19504+3318  & 21252 & 2019 Jan 12, 1.4\,h & 4956\\
J18010--3045  & 21253 & 2019 May 14, 3.0\,h & 4962\\
J18112--2641  & 21254 & 2019 Mar 6, 8.4\,h & 4955\\
J18013--3222  & 21255 & 2019 May 13, 10.4\,h & 5057\\ \hline
\end{tabular}
\end{minipage}
\end{table}

\begin{table}
\caption{Source Detection and Shifts Based on Optical/IR Register Matches\label{tab:shifts}}
\begin{minipage}{\linewidth}
\footnotesize
\begin{tabular}{ccccccccc} \hline \hline
IGR Name & ObsID & $N_{Chandra}$\footnote{The number of Chandra sources detected on the four ACIS-I detector chips.} & Survey & $N_{\rm matches}$\footnote{The number of matches between the Chandra detections and the survey catalog.} & $x_{\rm shift}$\footnote{The shifts in the x and y detector coordinate directions in pixels.  The conversion is 1 pixel = $0.492^{\prime\prime}$.} & $y_{\rm shift}$$^{c}$ & Avg. Residual\footnote{Average residual (in arcseconds) between the Chandra and O/IR sources.} & Max. Residual\footnote{Maximum residual (in arcseconds) between the Chandra and O/IR sources.}\\ \hline\hline
J18017--3542 & 20200 & 19 & Vista VVV & 5 & -0.08 & -0.20 & 0.55 & 0.90 \\
J12529--6351 & 20201 & 14 & Gaia EDR3 & 5 &  0.20 & -0.56 & 0.51 & 0.78 \\
J20084+3221  & 21248 & 11 & PanSTARRS & 2 &   --- &   --- &  --- &  --- \\
J18434--0508 & 21249 & 22 & Gaia EDR3 & 6 &  0.13 & -0.33 & 0.59 & 0.85 \\
J17040--4305 & 21250 & 12 & Gaia EDR3 & 2 &   --- &   --- &  --- &  --- \\
J19577+3339  & 21251 & 11 & Gaia EDR3 & 2 &   --- &   --- &  --- &  --- \\
J19504+3318  & 21252 & 13 & PanSTARRS & 2 &   --- &   --- &  --- &  --- \\
J18010--3045 & 21253 & 19 & Gaia EDR3 & 8 & -0.13 & -0.14 & 0.45 & 0.84 \\
J18112--2641 & 21254 & 18 & Vista VVV & 3 & -0.59 & -0.35 & 0.26 & 0.37 \\
J18013--3222 & 21255 & 15 & Gaia EDR3 & 7 & -0.27 &  0.49 & 0.57 & 0.72 \\ \hline
\end{tabular}
\end{minipage}
\end{table}

\begin{table}
\caption{Chandra Candidate Counterparts to IGR Sources\label{tab:positions}}
\begin{minipage}{\linewidth}
\footnotesize
\begin{tabular}{cccccccc} \hline \hline
IGR Name & CXOU Name & {\em Chandra} R.A. & {\em Chandra} Decl. & Uncertainty\footnote{90\% confidence.} & $\theta$\footnote{The angular distance between the center of the INTEGRAL error circle and the source.}/$\theta_{INTEGRAL}$\footnote{The size of the 90\% confidence INTEGRAL error radius given in \cite{krivonos17}.} & ACIS Counts\footnote{The number of counts, after background subtraction, measured by Chandra/ACIS-I in the 0.3--10\,keV band.} & Hardness\footnote{The hardness is given by $(C_{2}-C_{1})/(C_{2}+C_{1})$, where $C_{2}$ is the number of counts in the 2--10 keV band and $C_{1}$ is the number of counts in the 0.3--2 keV band.}\\
 & & (J2000) & (J2000) & (arcseconds) & (arcminutes) &  & \\ \hline \hline
J17040 & J170404.9--430537 & 17h04m04.92s & --43$^{\circ}$05$^{\prime}$37.9$^{\prime\prime}$ & 0.81 & 0.95/3.6 & $583\pm 25$ & +$0.72\pm 0.05$\\
J18010 & J180100.6--303958 & 18h01m00.61s & --30$^{\circ}$39$^{\prime}$58.4$^{\prime\prime}$ & 1.18 & 5.95/3.6 & $41\pm 8$ & +$0.48\pm 0.21$\\
J18017 & J180112.5--353912 & 18h01m12.53s & --35$^{\circ}$39$^{\prime}$12.1$^{\prime\prime}$ & 0.89 & 3.48/3.6 & $55\pm 8$ & +$0.86\pm 0.21$\\
J18112 & J181058.4--264115 & 18h10m58.41s & --26$^{\circ}$41$^{\prime}$15.7$^{\prime\prime}$ & 0.99 & 6.05/3.6 & $111\pm 12$ & +$0.43\pm 0.12$\\
J18434 & J184311.4--050545 & 18h43m11.43s & --05$^{\circ}$05$^{\prime}$45.2$^{\prime\prime}$ & 0.83 & 4.26/3.6 & $450\pm 22$ & +$0.48\pm 0.06$\\
J19504 & J195019.7+331416 & 19h50m19.73s & +33$^{\circ}$14$^{\prime}$16.7$^{\prime\prime}$ & 0.83 & 4.68/3.6 & $548\pm 24$ & +$0.19\pm 0.05$\\
J19577 & J195740.5+333828 & 19h57m40.59s & +33$^{\circ}$38$^{\prime}$28.2$^{\prime\prime}$ & 0.93 & 1.12/3.6 & $10\pm 4$ & +$0.80\pm 0.60$\\
J20084 & J200844.1+321818 & 20h08m44.14s & +32$^{\circ}$18$^{\prime}$18.3$^{\prime\prime}$ & 0.83 & 4.06/3.6 & $313\pm 19$ & +$0.73\pm 0.07$\\ \hline
\end{tabular}
\end{minipage}
\end{table}

\begin{table}
\caption{Chandra Spectral Parameters and Absolute Spurious Probabilities \label{tab:spectra_ch}}
\begin{minipage}{\linewidth}
\footnotesize
\begin{tabular}{ccccccccc} \hline \hline
IGR Name & $N_{\rm H}$\footnote{The errors on the parameters are 90\% confidence. The column density is calculated assuming \cite{wam00} abundances and \cite{vern96} cross sections.} & $N_{\rm H,Galactic}$\footnote{From the HI4PI survey \citep{bb16}.} & $\Gamma$ & Absorbed Flux\footnote{In units of erg\,cm$^{-2}$\,s$^{-1}$.} & Unabsorbed Flux\footnote{Only corrected for Galactic absorption.} & $C$/dof & $P_{\rm reject}$\footnote{The probability that an absorbed power-law does not provide a good description of the spectrum based on a calculation of the variance of $C$ according to the method described in \cite{kaastra17}.} & Probability\footnote{Absolute probability that a source of this brightness would be found by chance in the search region calculated using Equation 4.  The range comes from using the two $\log{N}-\log{S}$ distributions in Equations 2 and 3.}\\
  &  ($\times$$10^{22}$\,cm$^{-2}$) & ($\times$$10^{22}$\,cm$^{-2}$) & & (2--10\,keV) & (2--10\,keV) & & & \\ \hline \hline
J17040 & $0.5^{+0.7}_{-0.5}$ & 1.2 & $0.2\pm 0.4$ & $7.45\times 10^{-12}$ & $7.56\times 10^{-12}$ & 23/19\footnote{This includes a pileup correction.  There is an improvement to $C$/dof = 11/17 if an iron emission line is added (see Section 3).} & 41\% & 0.19--0.34\%\\
J18010 & $0.0^{+2.3}_{-0.0}$ & 0.3 & $0.4^{+1.2}_{-0.4}$ & $3.57\times 10^{-13}$ & $3.57\times 10^{-13}$ & 1.2/2 & ... & 10--21\%\\
J18017 & $5^{+12}_{-5}$ & 0.2 & $0.0^{+1.7}_{-1.3}$ & $7.05\times 10^{-13}$ & $7.08\times 10^{-13}$ & 2.2/3 & ... & 2.2--3.5\%\\ 
J18112 & $0.5^{+1.3}_{-0.5}$ & 0.3 & $0.9^{+0.7}_{-0.5}$ & $7.50\times 10^{-13}$ & $7.59\times 10^{-13}$ & 3.5/4 & ... & 5.8--8.9\%\\
J18434 & $0.4^{+0.5}_{-0.4}$ & 1.4 & $0.7\pm 0.3$ & $3.91\times 10^{-12}$ & $3.96\times 10^{-12}$ & 8/14 & ... & 0.59--0.80\%\\
J19504 & $1.1\pm 0.4$ & 0.5 & $1.8\pm 0.3$ & $2.17\times 10^{-12}$ & $2.24\times 10^{-12}$ & 22/18 & 41\% & 1.4--1.5\%\\
J19577 & $1.1^{+36}_{-1.1}$ & 0.9 & --$0.6^{+4.5}_{-1.5}$ & $2.52\times 10^{-13}$ & $2.57\times 10^{-13}$ & 0.2/1 & ... & 4.9-12\%\\
J20084 & $1.0^{+1.0}_{-0.9}$ & 0.9 & $0.1^{+0.4}_{-0.2}$ & $3.17\times 10^{-12}$ & $3.24\times 10^{-12}$ & 15/9\footnote{There is an improvement to $C$/dof = 7/7 if an iron emission line is added (see Section 3).} & 77\% & 0.69--0.86\%\\
\hline
\end{tabular}
\end{minipage}
\end{table}

\begin{table}
\caption{Gaia Identifications in EDR3\label{tab:gaia3}}
\begin{minipage}{\linewidth}
\scriptsize
\hspace{-1cm}
\begin{tabular}{cccccccccc} \hline \hline
IGR Name & Gaia Number & Separation\footnote{The angular separation between the Chandra position and the Gaia catalog position.} & $G$-Magnitude & $BP$-Magnitude & $RP$-Magnitude & Parallax & Astrometric Noise & Distance\footnote{From \cite{bj21}.} & Proper motion\\ 
 & (in EDR3) & (arcsec) &  &  &  & (milliarcsec) & (milliarcsec) & (kpc) & (milliarcsec/yr)\\\hline \hline
J17040 & 5965412985207709184 & 0.470 & $16.312\pm 0.005$ & $16.798\pm 0.014$ & $15.627\pm 0.012$ & $1.03\pm 0.06$ & 0 & $0.937\pm 0.046$ & $6.53\pm 0.07$\\
J18010 & 4044148421630416256 & 0.784 & $20.133\pm 0.034$ & --- & --- & --- & --- & --- & ---\\
J18017 & 4038975665929542784 & 0.273 & $15.714\pm 0.008$ & $15.827\pm 0.026$ & $15.451\pm 0.020$ & $0.61\pm 0.04$ & 0 & $1.58\pm 0.13$ & $6.89\pm 0.04$\\
J18112 & 4064533126752366464 & 0.790 & $16.625\pm 0.003$ & $16.991\pm 0.546$ & $15.429\pm 0.015$ & $0.37\pm 0.11$ & 0.673 & --- & $5.74\pm 0.09$\\
J18434 & 4256616815760182528 & 0.366 & $18.248\pm 0.011$ & $18.743\pm 0.043$ & $17.625\pm 0.045$ & $0.44\pm 0.15$ & 0 & $3.0^{+1.2}_{-0.9}$ & $3.98\pm 0.16$\\
J19504 & 2034764091072416384 & 0.460 & $18.698\pm 0.012$ & --- & --- & --$0.35\pm 0.16$ & 0.338 & --- & $0.12\pm 0.17$\\
J20084 & 2054890685756667392 & 0.250 & $19.461\pm 0.006$ & $20.893\pm 0.083$ & $18.223\pm 0.021$ & $0.15\pm 0.27$ & 0 & --- & $4.77\pm 0.23$\\ \hline
\end{tabular}
\end{minipage}
\end{table}

\begin{table}
\caption{VISTA VVV, 2MASS, UKIDSS, and PanSTARRS Identifications\label{tab:nir}}
\begin{minipage}{\linewidth}
\scriptsize
\begin{tabular}{ccccccccc} \hline \hline
IGR Name & Catalog & Source (arcsec) & Separation\footnote{The angular separation between the {\em Chandra} position and the catalog position.} & $Y$ & $J$ & $H$ & $K$/$K_{s}$ & Class\footnote{For VISTA VVV and UKIDSS, the classification is based on the spatial profile, where --2 is a probable star, --1 is a star with probability $\ge$90\%, and 1 is a galaxy with probability $\ge$90\%.  For PanSTARRS, quality flag 61 indicates that the source is extended, and flags 52 and 60 indicate that the criterion for the detection of source extension is not met.}\\\hline \hline
J17040 & VISTA VVV & VVV J170404.96--430538.07 & 0.530 & $14.885\pm 0.003$ & $14.627\pm 0.003$ & $14.499\pm 0.006$ & $K_{s}=14.197\pm 0.001$ & --1\\ \hline
J18010 & VISTA VVV & VVV J180100.64--303957.61 & 0.879 & -- & $18.404\pm 0.487$ & $16.681\pm 0.173$ & $K_{s}=15.640\pm 0.095$ & --1\\ \hline
J18017 & VISTA VVV & VVV J180112.51--353912.10 & 0.191 & $15.288\pm 0.011$ & $15.008\pm 0.008$ & $14.945\pm 0.014$ & $K_{s}=14.821\pm 0.022$ & --1\\ \hline
J18112 & VISTA VVV & VVV J181058.38--264114.93 & 0.839 & $14.467\pm 0.008$ & $13.943\pm 0.008$ & $13.290\pm 0.007$ & $K_{s}=13.024\pm 0.008$ & --1\\
J18112 & PanSTARRS & 75972727431145261 & 0.776 & $15.425\pm 0.002$ & -- & -- & -- & 52\\ \hline
J18434 & UKIDSS &     J184311.43--050545.6 & 0.403 & -- & $18.932\pm 0.173$ & $18.087\pm 0.206$ & $16.883\pm 0.131$ & --1\\
J18434 & PanSTARRS & 101882807975695145 & 0.364 & $17.671\pm 0.027$ & -- & -- & -- & 52\\ \hline
J19504 & 2MASS &         19501973+3314166 & 0.115 & -- & $15.423\pm 0.060$ & $14.542\pm 0.068$ & $13.682\pm 0.049$ & --\\
J19504 & PanSTARRS & 147882975821426046 & 0.166 & $17.136\pm 0.026$ & -- & -- & -- & 61\\ \hline
J20084 & UKIDSS &     J200844.16+321818.2 & 0.375 & -- & $16.037\pm 0.007$ & $15.306\pm 0.006$ & $14.885\pm 0.011$ & --1\\
J20084 & PanSTARRS & 146763021840106679 & 0.279 & $17.652\pm 0.033$ & -- & -- & -- & 60\\ \hline
\end{tabular}
\end{minipage}
\end{table}

\begin{table}
\caption{Galactic Sources with Distances\label{tab:distances}}
\begin{minipage}{\linewidth}
\footnotesize
\begin{tabular}{cccc} \hline \hline
Parameter & J17040 & J18017 & J18434\\ \hline
$L_{x}$\footnote{2--10\,keV luminosity in erg\,s$^{-1}$ with 68\% confidence errors.} & $(7.9\pm 0.8)\times 10^{32}$ & $(2.1\pm 0.4)\times 10^{32}$ & $(4.2^{+3.4}_{-2.6})\times 10^{33}$\\
$E(B-V)$ & $0.237^{+0.010}_{-0.012}$ & $0.326^{+0.005}_{-0.016}$ & $0.75^{+0.09}_{-0.17}$\\
$A_{V}$ & $0.73^{+0.03}_{-0.04}$ & $1.01^{+0.02}_{-0.05}$ & $2.3^{+0.3}_{-0.5}$\\ \hline
\multicolumn{4}{c}{Near-IR analysis}\\ \hline
$A_{J}$ & $0.190^{+0.008}_{-0.010}$ & $0.262^{+0.005}_{-0.013}$ & $0.60^{+0.08}_{-0.13}$\\
$A_{J}$--$A_{H}$ & $0.066^{+0.003}_{-0.004}$ & $0.091^{+0.002}_{-0.005}$ & $0.21^{+0.03}_{-0.05}$\\
$J$--$H$--($A_{J}$--$A_{H}$) & $0.062^{+0.008}_{-0.007}$ & --$0.028^{+0.017}_{-0.016}$ & $0.64\pm 0.27$\\
$T_{\rm eff}$ (K) & $7680\pm 90$ & $9600\pm 800$ & 2200--5300\\
$M_{J}$ & $4.58\pm 0.11$ & $3.75\pm 0.18$ & $5.9^{+0.9}_{-0.7}$\\
Limit on type\footnote{Spectral type if the $J$-band luminosity is dominated by emission from a star.  If there is another contribution (from an accretion disk, for example), then any stellar component in the system would have a later spectral type.} & K2V or later & G2V or later & K5V or later\\
$T_{\rm type}$ (K) & $<$5100 & $<$5770 & $<$4400\\ \hline
\multicolumn{4}{c}{Gaia analysis}\\ \hline
$A_{G}$ & $0.66\pm 0.03$ & $0.910^{+0.014}_{-0.045}$ & $2.1^{+0.3}_{-0.5}$\\
$A_{BP}$--$A_{RP}$ & $0.37\pm 0.02$ & $0.516^{+0.008}_{-0.026}$ & $1.17^{+0.15}_{-0.26}$\\
$BP$--$RP$--($A_{BP}$--$A_{RP}$) & $0.79\pm 0.03$ & --$0.14^{+0.04}_{-0.03}$ & --$0.05^{+0.27}_{-0.16}$\\
$T_{\rm eff}$ (K) & $5880\pm 110$ & $>$11,000 & $>$7900\\
$M_{G}$ & $5.79\pm 0.11$ & $3.81\pm 0.18$ & $3.8^{+1.0}_{-0.8}$\\
Limit on type & K1V or later & F7.5V or later & F3V or later\\
$T_{\rm type}$ (K) & $<$5170 & $<$6200 & $<$6750\\ \hline\hline
\end{tabular}
\end{minipage}
\end{table}

\begin{table}
\caption{Summary of Results\label{tab:summary}}
\begin{minipage}{\linewidth}
\footnotesize
\begin{tabular}{ccccl} \hline \hline
IGR Name & Chandra counterpart or & Galactic or Extragalactic & Source Type & Evidence\\
         & 2--10\,keV flux limit  &     &        &         \\ \hline
J12529--6351\footnote{The Chandra limit indicates that the IGR source is transient, highly variable, or highly absorbed.} & $<$$1\times 10^{-13}$ & --- & --- & ---\\
J17040--4305 & CXOU J170404.9--430537 & Galactic & magnetic CV? & Gaia distance ($0.937\pm 0.046$\,kpc)\\
J18010--3045 & CXOU J180100.6--303958? & ? & ? & ?\\
J18013--3222$^{a}$ & $<$$1\times 10^{-13}$ & --- & --- & ---\\
J18017--3542 & CXOU J180112.5--353912  & Galactic & magnetic CV? & Gaia distance ($1.58\pm 0.13$\,kpc)\\
J18112--2641 & CXOU J181058.4--264115  & Galactic & magnetic CV? & Gaia proper motion; variability in $Y$\\
J18434--0508 & CXOU J184311.4--050545 & Galactic & magnetic CV? & Gaia distance ($3.0^{+1.2}_{-0.9}$\,kpc)\\
J19504+3318 & CXOU J195019.7+331416 & Extragalactic? & AGN? & WISE colors\\
J19577+3339 & CXOU J195740.5+333828 & Extragalactic? & AGN? & Bright radio source\\
J20084+3221 & CXOU J200844.1+321818 & Galactic & magnetic CV? or HMXB? & Gaia proper motion\\ \hline\hline
\end{tabular}
\end{minipage}
\end{table}



\end{document}